\RequirePackage{filecontents}

\RequirePackage{snapshot}
\documentclass[aps,pra,reprint,showpacs,secnumarabic]{preprint}

\renewcommand\documentclass[2][]{}
\let\origparagraph=\paragraph
\AtBeginDocument{%
  \let\paragraph=\origparagraph
}
\makeatother
\documentclass[reprint,aps,pra,showpacs,floatfix]{revtex4-1}

\usepackage[T1]{fontenc}
\usepackage{mathtools}
\usepackage{textcomp}
\usepackage{txfonts}
\usepackage{tgtermes}
\usepackage[altg]{qtxmath}
\usepackage{MnSymbol}
\makeatletter
\newcommand{\raisemath}[1]{\mathpalette{\raisem@th{#1}}}
\newcommand{\raisem@th}[3]{\raisebox{#1}{$#2#3$}}
\def\hbar{\ensuremath{\mathrlap{\raisemath{-0.1ex}{{\mathchar'26}}}\mkern-1mu h}}
\makeatother
\renewcommand*{\vec}[1]{\boldsymbol{#1}}

\usepackage{microtype}
\usepackage{graphicx}
\usepackage{braket}
\graphicspath{{figures/}}
\usepackage[breaklinks=true,
pdfauthor={Rico Erhard, Heiko Bauke},
pdftitle={Spin effects in Kapitza-Dirac scattering at light with elliptical polarization}]{hyperref}

\newcommand*{\D}{\mathrm{d}}
\newcommand*{\e}{\mathrm{e}}
\newcommand*{\I}{\mathrm{i}}
\newcommand*{\trans}[1]{#1^{\mathsf{T}}}

\newcommand*{\corot}{^{\lcurvearrowup\lcurvearrowup}}
\newcommand*{\antirot}{^{\lcurvearrowup\rcurvearrowdown}}

\newcommand*{\commut}[2]{\mathchoice{\left[#1,#2\right]}{[#1,#2]}{[#1,#2]}{[#1,#2]}}
\newcommand*{\acommut}[2]{\mathchoice{\left\{#1,#2\right\}}{\{#1,#2\}}{\{#1,#2\}}{\{#1,#2\}}}

\newcommand*{\E}{\mathcal{E}}

\begin{document}

\title{Spin effects in Kapitza-Dirac scattering at light with elliptical polarization}

\author{Rico Erhard}
\affiliation{Max-Planck-Institut f{\"u}r Kernphysik, Saupfercheckweg~1, 69117~Heidelberg, Germany}
\author{Heiko Bauke}
\email{heiko.bauke@mpi-hd.mpg.de}
\affiliation{Max-Planck-Institut f{\"u}r Kernphysik, Saupfercheckweg~1, 69117~Heidelberg, Germany}

\begin{abstract}
  The Kapitza-Dirac effect, which refers to electron scattering at
  standing light waves, is studied in the Bragg regime with
  counterpropagating elliptically polarized electromagnetic waves with
  the same intensity, wavelength, and degree of polarization for two
  different setups.  In the first setup, where the electric field
  components of the counterpropagating waves have the same sense of
  rotation, we find distinct spin effects.  The spins of the scattered
  electrons and of the nonscattered electrons, respectively, precess
  with a frequency that is of the order of the Bragg-reflection Rabi
  frequency.  When the electric-field components of the
  counterpropagating waves have an opposite sense of rotation, which is
  the second considered setup, the standing wave has linear polarization,
  and no spin effects can be observed.  Our results are based on
  numerical solutions of the time-dependent Dirac equation and the
  analytical solution of a relativistic Pauli equation, which accounts
  for the leading relativistic effects.
\end{abstract}


\maketitle

\allowdisplaybreaks

\section{Introduction}
\label{sec:intro}

Kapitza-Dirac scattering
\cite{Kapitza:Dirac:1933:reflection_of,Batelaan:2000:Kapitza-Dirac_effect},
which is the scattering of massive particles on a standing light wave,
emerges as a remarkable consequence of the wave-particle duality of
quantum objects.  When Kapitza and Dirac first predicted electron
scattering at a standing light wave in 1933, an experimental realization
was out of reach.  The advent of high-intensity laser facilities,
however, rendered such an experimental realization possible.  The first
clear observations of the Kapitza-Dirac effect in the so-called Bragg
regime were achieved by employing atoms
\cite{Adams:Sigel:etal:1994:Atom_optics,Freyberger:Herkommer:etal:1999:Atom_Optics}
in 1986 \cite{Gould:Ruff:etal:1986:Diffraction_of} and in the so-called
diffraction regime \cite{Martin:Oldaker:etal:1988:Bragg_scattering} in
1988.  Recently, strong-field Kapitza-Dirac scattering of neutral atoms
was realized through frustrated tunneling ionization leading to an
extreme acceleration of the employed helium atoms
\cite{Eilzer:Zimmermann:etal:2014:Strong-Field_Kapitza-Dirac}.  The
Kapitza-Dirac effect with electrons was first achieved in 1988
\cite{Bucksbaum:Schumacher:etal:1988:High-Intensity_Kapitza-Dirac} in
the diffraction regime.  The scattering of electrons in the Bragg regime
was demonstrated experimentally in 2001
\cite{Freimund:Aflatooni:Batelaan:2001:Observation_of,Freimund:Batelaan:2002:Bragg_Scattering},
which comes closest to the diffraction process as proposed originally by
Kapitza and Dirac \cite{Kapitza:Dirac:1933:reflection_of}.  These
experiments stimulated also renewed theoretical interest in this effect
\cite{Efremov:Fedorov:2000:Wavepacket_theory,Li:Zhang:etal:2004:Theory_of,Batelaan:2007:Colloquium_Illuminating}
and generalizations thereof.  For example, in
Refs.~\cite{Sancho:2010:Two-particle_Kapitza-Dirac,Sancho:2011:Bragg_regime}
two-particle Kapitza-Dirac scattering was studied, and also
Kapitza-Dirac scattering of electrons from a bichromatic standing light
wave with linear polarization
\cite{Dellweg:Muller:2015:Kapitza-Dirac_scattering} or circular
polarization \cite{McGregor:Huang:etal:2015:Spin-dependent_two-color}
has been considered recently.  In dielectric media, electrons may also
diffract from traveling light waves
\cite{Hayrapetyan:Grigoryan:etal:2015:Kapitza-Dirac_effect}.

As the electron couples to electromagnetic fields not only via its
charge but also via its spin degree of freedom, it is natural to ask if
electron-spin dynamics may be observed in the Kapitza-Dirac effect
\cite{Freimund:Aflatooni:Batelaan:2001:Observation_of}.  Freimund and
Batelaan found a vanishingly small spin-flip probability in the
investigated laser setup and parameter regime by simulating the
Bargmann-Michel-Telegdi equations
\cite{Freimund:Batelaan:2003:Stern-Gerlach}.  Another derivation
\cite{Rosenberg:2004:Extended_theory} solved the Pauli equation
perturbatively via second-quantization methods in the diffraction regime
but only found tiny spin effects for a setup with linearly polarized
light waves.  Spin effects, however, can be observed in the
Kapitza-Dirac effect if the electron enters the laser field with a
certain relativistic initial momentum, which must fulfill a spin-flip
resonance condition
\cite{Ahrens:Bauke:etal:2012:Spin_Dynamics,Ahrens:Bauke:etal:2013:Kapitza-Dirac_effect},
or in specific laser setups with circular polarization
\cite{McGregor:Huang:etal:2015:Spin-dependent_two-color}.

As a result of emerging novel high-intensity light sources such as the
Extreme Light Infrastructure Ultra High Field Facility, which envisage
to provide field intensities in excess of $10^{20}\,\mathrm{W/cm^2}$ and
field frequencies in the \mbox{x-ray} domain
\cite{Mourou:Fisch:etal:2012:Exawatt-Zettawatt_pulse}, spin effects in
relativistic light-matter interaction became of wide interest
\cite{Hu:Keitel:1999:Spin_Signatures,Walser:Urbach:etal:2002:Spin_and,
  Faisal:Bhattacharyya:2004:Spin_Asymmetry,
  Brodin:Marklund:etal:2011:Spin_and,
  Skoromnik:Feranchuk:etal:2013:Collapse-and-revival_dynamics}. In
particular, the interaction between the electron's spin and the degree
of elliptical polarization of an external strong electromagnetic field
has been considered.  For example, it was recently shown that electrons
scattering nonresonantly at standing light waves of counterpropagating
plane waves with elliptical polarization can exhibit spin precession
\cite{Bauke:Ahrens:etal:2014:Electron-spin_dynamics,Bauke:Ahrens:etal:2014:Electron-spin_dynamics_long}.
Strong spin effects have also been found in electron-positron pair
production in strong electromagnetic fields with elliptical polarization
\cite{Wollert:Bauke:etal:2015:Spin_polarized,Muller:Muller:2012:Longitudinal_spin}
as well as in strong-field ionization via electromagnetic fields with
elliptical polarization
\cite{Barth:Smirnova:2014:Hole_dynamics,Yakaboylu:Klaiber:etal:2015:Above-threshold_ionization}.

In this paper we will study the Kapitza-Dirac effect with electrons in
standing light waves of counterpropagating plane waves with elliptical
polarization and especially study possible spin effects.  This article
is organized as follows: In Sec.~\ref{sec:setup} we will specify the
considered configurations of the time-dependent electromagnetic fields.
The effect of the fast oscillating fields can be described by
time-independent ponderomotive potentials, which will be derived in
Sec.~\ref{sec:ponderomotive_pot}. The resulting time evolution by the
ponderomotive potentials will be determined analytically and compared to
a numerical solution of the time-dependent Dirac equation in
Sec.~\ref{sec:solution}.  A summary and our conclusions follow in
Sec.~\ref{sec:conclusions}.

\section{Electrons in counterpropagating electromagnetic waves with
  elliptical polarization}
\label{sec:setup}

\begin{figure}
  \centering
  \includegraphics[scale=0.38]{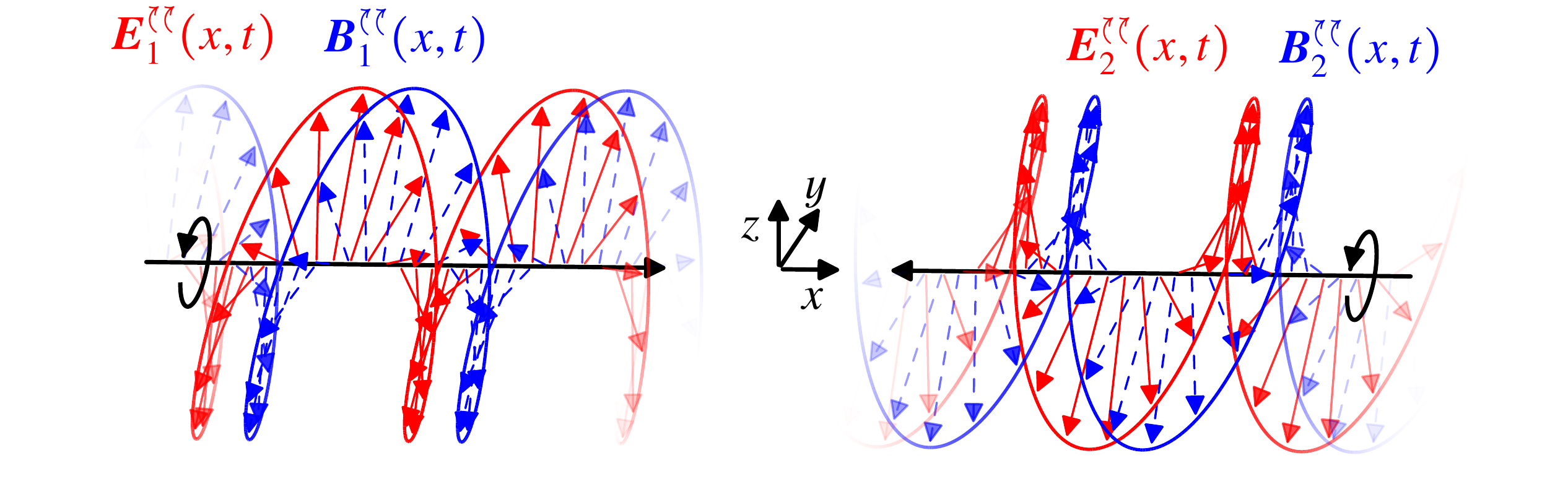}\\[4ex]
  \includegraphics[scale=0.38]{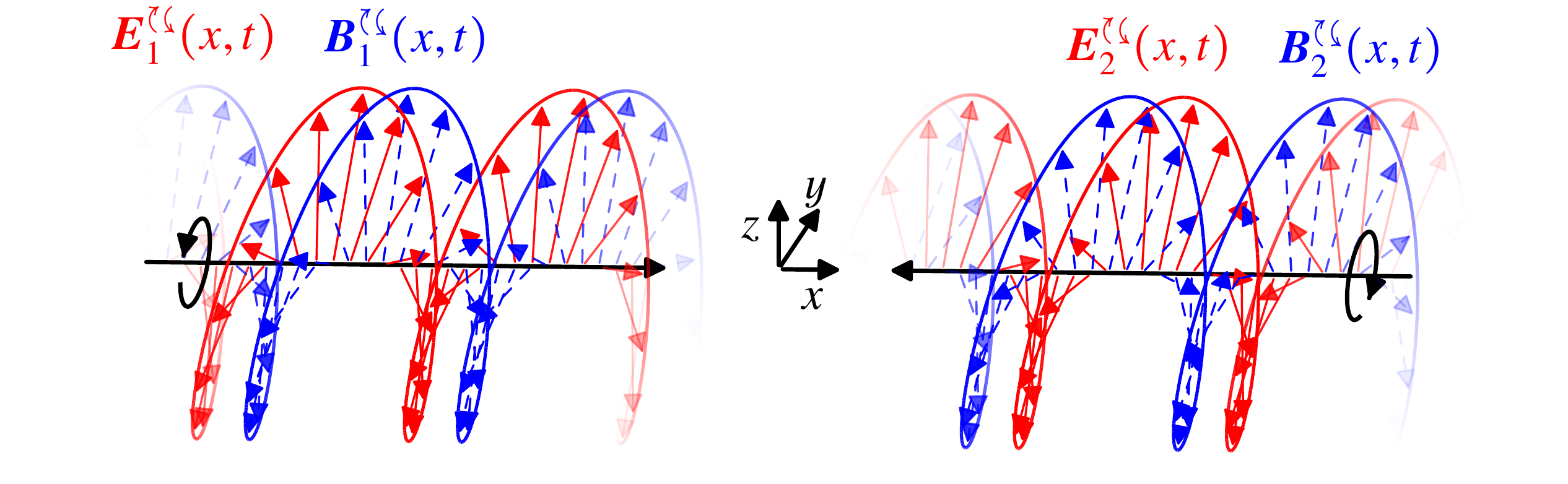}
  \caption{(Color online) Schematic illustration of the two considered
    field configurations.  Two electromagnetic plane waves with
    elliptical polarization traveling into opposite directions are
    superimposed.  The sense of rotation of the electric field vector
    may be (top) equal (corotating setup) or (bottom) opposite
    (antirotating setup).  The sense of rotation and the propagation
    direction are indicated by black arrows, red solid arrows represent
    electric-field components, and blue dashed arrows represent
    magnetic-field components.}
  \label{fig:waves}
\end{figure}

Electron scattering at a standing light wave, which is composed of two
counterpropagating laser waves with equal wavelength $\lambda$, equal
electric field amplitude $\hat E$, and the same degree of ellipticity,
will be considered.  Depending on the relative sense of rotation of the
electric-field component of the two plane-wave fields traveling along
the $x$~axis, the electric- and magnetic-field components are given by
\pagebreak[1]
\begin{subequations}
  \begin{align}
    \label{eq:E_co}
    \vec E\corot_{1,2}(x, t) & = \hat E
    \begin{pmatrix}
      0 \\ \cos(kx \mp \omega t) \\ \cos(kx\mp\omega t \pm \eta)
    \end{pmatrix} \,,\\
    \label{eq:B_co}
    \vec B\corot_{1,2}(x, t) & = \frac{\hat E}{c}
    \begin{pmatrix}
      0 \\ \mp\cos(kx \mp \omega t \pm \eta) \\ \pm\cos(kx\mp\omega t)
    \end{pmatrix} \,,
  \end{align}
\end{subequations}
or\pagebreak[1]
\begin{subequations}
  \begin{align}
    \label{eq:E_anti}
    \vec E\antirot_{1,2}(x, t) & = \hat E
    \begin{pmatrix}
      0 \\ \cos(kx \mp \omega t) \\ \cos(kx\mp\omega t + \eta)
    \end{pmatrix} \,,\\
    \label{eq:B_anti}
    \vec B\antirot_{1,2}(x, t) & = \frac{\hat E}{c}
    \begin{pmatrix}
      0 \\ \mp\cos(kx \mp \omega t + \eta) \\ \pm\cos(kx\mp\omega t)
    \end{pmatrix} \,,
  \end{align}
\end{subequations}
respectively, where $c$ denotes the speed of light, $k=2\pi/\lambda$ is
the wave number, and $\omega=2\pi c/\lambda$.  The parameter
$\eta\in(-\pi,\pi]$ determines the degree of ellipticity, with $\eta=0$
and $\eta=\pi$ corresponding to linear polarization and $\eta=\pm\pi/2$
corresponding to circular polarization.  The two electromagnetic waves
given by \eqref{eq:E_co} and \eqref{eq:B_co} rotate in parallel
(corotating setup), whereas the two electromagnetic waves given by
\eqref{eq:E_anti} and \eqref{eq:B_anti} rotate opposite to each other
(antirotating setup); see Fig.~\ref{fig:waves} for an illustration.  The
total electromagnetic fields and the corresponding vector potentials of
these setups follow as
\begin{subequations}
  \begin{align}
    \vec E\corot(x, t) & = 2\hat E\cos(kx)
    \begin{pmatrix}
      0 \\ \cos(\omega t) \\ \cos(\omega t - \eta)
    \end{pmatrix} \,, \\
    \vec B\corot(x, t) & = \frac{2\hat E}{c}\sin(kx)
    \begin{pmatrix}
      0 \\ -\sin(\omega t - \eta) \\ \sin(\omega t) 
    \end{pmatrix} \,, \\
    \label{eq:A_co}
    \vec A\corot(x, t) & = -\frac{2\hat E}{\omega}\cos(kx)
    \begin{pmatrix}
      0 \\ \sin(\omega t) \\ \sin(\omega t - \eta)
    \end{pmatrix}  
  \end{align}
\end{subequations}
for the corotating setup and 
\begin{subequations}
  \begin{align}
    \vec E\antirot(x, t) & = 2\hat E\cos(\omega t)
    \begin{pmatrix}
      0 \\ \cos(kx) \\ \cos(kx + \eta)
    \end{pmatrix} \,, \\
    \vec B\antirot(x, t) & = \frac{2\hat E}{c}\sin(\omega t)
    \begin{pmatrix}
      0 \\ -\sin(kx + \eta) \\ \sin(kx) 
    \end{pmatrix} \,, \\
    \label{eq:A_anti}
    \vec A\antirot(x, t) & = -\frac{2\hat E}{\omega}\sin(\omega t)
    \begin{pmatrix}
      0 \\ \cos(kx) \\ \cos(kx + \eta)
    \end{pmatrix} 
  \end{align}
\end{subequations}
for the antirotating setup, respectively. 

The quantum-mechanical evolution of an electron of mass $m$ and charge
$q=-e$ in the laser field with the vector potential \eqref{eq:A_co} or
\eqref{eq:A_anti} is governed by the quasi-one-dimensional Dirac
equation
\begin{equation}
  \label{eq:Dirac_equation}
  \I \hbar\dot \Psi(x, t)  = 
  \left(
    -c \I\hbar \alpha_x\frac{\partial}{\partial x}
    -c q w(t) \vec \alpha\cdot \vec A(x, t) + m c^2 \beta
  \right) \Psi(x, t)
\end{equation}
with the Dirac matrices
$\vec\alpha=\trans{(\alpha_x,\alpha_y,\alpha_z)}$ and $\beta$
\cite{Gross:2004:Relativistic_Quantum,Thaller:2005:Advanced_Visual} and
the reduced Planck constant $\hbar$.  This form of the Dirac equation
can be justified by starting from the fully three-dimensional Dirac
equation and noting that the canonical momentum is preserved in the $y$
and in $z$~directions due to the special form of the vector potentials
\eqref{eq:A_co} and \eqref{eq:A_anti}.  Thus, the three-dimensional
Dirac equation separates into three one-dimensional equations, of which
one is given by \eqref{eq:Dirac_equation} and the remaining two describe
free motion in the $y$ and in $z$~directions.  In
\eqref{eq:Dirac_equation} the window function
\begin{equation}
  \label{eq:turn_on_and_off}
  w(t) = 
  \begin{cases}
    \sin^2 \frac{\pi t}{2\Delta T} & \text{if $0\le t\le \Delta T$,} \\
    1 & \text{if $\Delta T\le t\le T-\Delta T$,} \\
    \sin^2 \frac{\pi (T-t)}{2\Delta T} & \text{if $T-\Delta T\le t\le T$} 
  \end{cases}
\end{equation}
has been introduced, which allows for a smooth turn-on and turn-off of
the laser field.

In Sec.~\ref{sec:solution} we will present numerical solutions of the
time-dependent Dirac equation \eqref{eq:Dirac_equation} which are
obtained solving this equation on a one-dimensional grid of size
$\lambda$ with periodic boundary conditions by a Fourier split-operator
method
\cite{Braun:Su:etal:1999:Numerical_approach,Bauke:Keitel:2011:Accelerating_Fourier}.
Due to the periodic boundary condition the wave function can be expanded
into a discrete set of momentum eigenstates
\begin{equation}
  \label{eq:Dirac_momentum_eigen}
  \psi_n^{\gamma}(x) = 
  \sqrt{\frac{k}{2 \pi}} \,u_n^\gamma \e^{\I n k x}\,,
\end{equation}
which are simultaneous eigenstates of the free Dirac Hamiltonian, the
$x$ component of the canonical momentum operator (with eigenvalue $n
k\hbar$), and the $z$~component of the Foldy-Wouthuysen spin operator
\cite{Foldy:Wouthuysen:1950:On_Dirac}.  The upper index
$\gamma\in\{+{\uparrow}, -{\uparrow}, +{\downarrow}, -{\downarrow}\}$ in
$\psi_n^{\gamma}(x)$ indicates the spin state, and the sign of the
energy eigenvalue and $u_n^\gamma$ is defined as
\begin{subequations}
  \label{eq:Dirac_bi-spinors}
  \begin{align}
    u_n^{+\uparrow/+\downarrow} & = \sqrt{\frac{\E_n + m c^2}{2 \E_n}}
    \begin{pmatrix}
      \vec\chi^{\uparrow/\downarrow} \\[0.67ex]
      \frac{n c k \hbar\sigma_x}{\E_n + m c^2} \vec\chi^{\uparrow/\downarrow}
    \end{pmatrix} \,,\\
    u_n^{-\uparrow/-\downarrow} & = \sqrt{\frac{\E_n + m c^2}{2 \E_n}}
    \begin{pmatrix}
      - \frac{n c k \hbar\sigma_x}{\E_n + m c^2} \vec\chi^{\uparrow/\downarrow} \\[0.67ex]
      \vec\chi^{\uparrow/\downarrow}
    \end{pmatrix} \,,
  \end{align}
\end{subequations}
with $\vec\chi^\uparrow = \trans{(1, 0)}$ and $\vec\chi^\downarrow =
\trans{(0, 1)}$ and the relativistic energy-momentum relation
\begin{equation}
  \label{eq:relativistic_enery_momentum_relation_modes}
  \E_n = \sqrt{(m c^2)^2 + (n c k\hbar)^2} \,.
\end{equation}
In the so-called Bragg regime of the Kapitza-Dirac effect
\cite{Batelaan:2000:Kapitza-Dirac_effect,Batelaan:2007:Colloquium_Illuminating}
the electron has to meet the resonance condition, which requires that
the electron's momentum in the propagation direction of the light field
equals $\pm\hbar k$.  Thus, we will employ $\psi_{-1}^{+{\uparrow}}(x)$
as a canonical initial quantum state.

\section{Ponderomotive potentials}
\label{sec:ponderomotive_pot}

The Bragg regime of the Kapitza-Dirac effect is realized for
sufficiently weak electromagnetic fields, more precisely, if the energy
that is gained in a single laser cycle, which is of the order
$c|q|\hat E/\omega$, is small compared to the energy of a single photon,
which is $\hbar\omega$.  Furthermore, we will assume that the electron's
kinetic energy remains small compared to its rest mass energy $mc^2$ for
all times.  Thus, $c|q|\hat E/\omega\ll \hbar\omega<mc^2$.  As a consequence
of the Bragg condition, nonrelativistic initial momentum implies
$\lambda\gg \lambda_\mathrm{C}=\hbar/(mc)$, where $\lambda_\mathrm{C}$
is the electron's reduced Compton wavelength.  The nonrelativistic
electron momentum justifies the application of a weakly relativistic
theory.  The weakly relativistic limit of the Dirac equation may be
reached via a Foldy-Wouthuysen transformation
\cite{Foldy:Wouthuysen:1950:On_Dirac,Frohlich:Studer:1993:Gauge_invariance},
which yields, for a now two-component wave function $\Psi(\vec r, t)$,
\begin{multline}
  \label{eq:Dirac_FW_equation}
  \I \hbar\dot \Psi(\vec r, t)  = 
  \Bigg( 
  \frac{( -\I\hbar\vec\nabla - q \vec A(\vec r, t) )^2}{2m} -
  \frac{q\hbar}{2m}\vec\sigma\cdot \vec B(\vec r, t) 
  \\
  -
  \frac{( -\I\hbar\vec\nabla - q \vec A(\vec r, t) )^4}{8m^3c^2} -
  \frac{q^2\hbar^2}{8m^3c^4}(c^2\vec B(\vec r, t)^2-\vec E(\vec r, t)^2) \\
  -
  \frac{q\hbar}{4m^2c^2 }\vec\sigma\cdot (\vec E(\vec r, t) \times (
  -\I\hbar\vec\nabla - q\vec A(\vec r, t) ) ) -
  \frac{q\hbar^2}{8m^2c^2}\vec\nabla\cdot\vec{E}(\vec r, t) \\+
  \frac{q\hbar}{8m^3c^2}\acommut{\vec\sigma\cdot\vec B(\vec r, t)}{( -\I\hbar\vec\nabla - q \vec A(\vec r, t) )^2}
  \Bigg)\,\Psi(\vec r, t)\,,
\end{multline}
with the magnetic vector potential $\vec A(\vec r, t)$ and $\vec B(\vec
r, t)=\mbox{$\vec\nabla\times\vec A(\vec r, t)$}$, $\vec E(\vec r,
t)=-\dot{\vec A}(\vec r, t)$, and the vector of Pauli matrices
$\vec\sigma=\trans{(\sigma_x,\sigma_y,\sigma_z)}$.  Taking into account
that the electron's kinematic momentum remains small compared to $mc$
and that $\lambda\gg \lambda_\mathrm{C}=\hbar/(mc)$ allows us to neglect
various terms in Eq.~\eqref{eq:Dirac_FW_equation}, and the specific form of
the vector potentials \eqref{eq:A_co} and \eqref{eq:A_anti} leads
finally to \allowdisplaybreaks[0]\pagebreak[2]
\begin{multline}
  \label{eq:Dirac_FW_equation_1d}
  \I \hbar\dot \Psi(x, t)  = 
  \Bigg( 
   -\frac{\hbar^2}{2m}\frac{\partial^2}{\partial x^2} + \frac{q^2}{2m} \vec A(x, t)^2 -
   \frac{q\hbar}{2m}\vec\sigma\cdot \vec B(x, t) 
   \\ 
   -\frac{q\hbar}{4m^2c^2 }\vec\sigma\cdot (\vec E(x, t) \times (
   -\I\hbar\vec\nabla_x - q\vec A(x, t) ) ) 
   \Bigg)\,\Psi(x, t) \,,
\end{multline}
\allowdisplaybreaks%
where $\vec\nabla_x$ denotes the differential operator
$\vec\nabla_x=\trans{(\partial_x, 0, 0)}$.  This is the nonrelativistic
Pauli equation, which is amended by a relativistic correction due to
spin-orbit coupling \mbox{($\sim \vec E \times(-\I\hbar\nabla_x)$)} and
the coupling of the electron's spin to the photonic spin density of the
electromagnetic wave \mbox{($\sim \vec E \times\vec A$)}
\cite{Bauke:Ahrens:etal:2014:Electron-spin_dynamics,Bauke:Ahrens:etal:2014:Electron-spin_dynamics_long}.
It incorporates the dominating terms of \eqref{eq:Dirac_FW_equation}
that account for the electron's spatial motion as well as for the spin
dynamics.

As shown in \cite{Batelaan:2007:Colloquium_Illuminating} for the case of
linear polarization, the Kapitza-Dirac effect can be modeled by a
time-independent Hamiltonian with suitable ponderomotive potentials.
Thus, it appears attractive to replace the time-dependent Hamiltonian in
Eq.~\eqref{eq:Dirac_FW_equation_1d} by a time-independent Hamiltonian,
which we will derive in the following by means of a Magnus expansion
\cite{Magnus:1954:On_exponential,Blanes:Casas:etal:2009:Magnus_expansion}.
The Magnus expansion gives an exponential representation of the solution
of a first-order homogeneous linear equation of the type
\begin{equation}
  \dot f(t) = \tilde H(t) f(t) 
\end{equation}
in the form
\begin{equation}
  \label{eq:Magnus}
  f(t) = \exp( \tilde U_1(t) + \tilde U_2(t) + \dots) f(0) \,.
\end{equation}
The main advantage of the Magnus expansion is that each truncated series
\eqref{eq:Magnus} of any order preserves the unitary character of the
quantum-mechanical time evolution.  Explicitly, the first two terms of
the series \eqref{eq:Magnus} are
\begin{equation}
  \tilde U_1(t) = 
  \int_0^t \tilde H(t_1)\,\D t_1
\end{equation}
and
\begin{equation}
  \tilde U_2(t) = 
  \frac{1}{2}\int_0^t\int_0^{t_1} \commut{\tilde H(t_1)}{\tilde
    H(t_2)}\,\D t_2\,\D t_1 \,.
\end{equation}
The term $\tilde U_2(t)$ and higher-order corrections account for the
fact that $\tilde H(t)$ may not commute with itself at different times
$t$ and in this way implement time ordering.

Applying the Magnus expansion in second order to
\eqref{eq:Dirac_FW_equation_1d} with the vector potential of the
corotating case \eqref{eq:A_co} and the corresponding electromagnetic
fields yields
\begin{widetext}
  \begin{equation}
    \label{eq:U_co}
    \Psi(x, t) = \exp\Bigg(
      -\frac{\I t}{\hbar}\Bigg(
        -\frac{\hbar^2}{2m}\frac{\partial^2}{\partial x^2} + 
        \frac{2q^2\hat E^2}{m\omega^2}\cos^2(kx) -
        \frac{\hbar q^2\hat E^2}{m^2c^2\omega}\sin\eta\,(\sin^2(kx)-\cos^2(kx))\,\sigma_x 
        + \dots
      \Bigg) 
    \Bigg)\Psi(x, 0) \,,
  \end{equation}
  where we have neglected in the exponent terms which do not grow
  linearly with time and/or are small (compared to the included terms)
  in the Bragg regime, i.\,e., $|q|\hat E/(\omega mc)\ll 1$ with
  $\lambda\gg \lambda_\mathrm{C}=\hbar/(mc)$.  Equation \eqref{eq:U_co}
  represents the time-evolution operator of the equation\pagebreak[1]
  \begin{equation}
    \label{eq:H_pond_co}
    \I \hbar\dot \Psi(x, t)  = 
    \left( 
      -\frac{\hbar^2}{2m}\frac{\partial^2}{\partial x^2} + 
      \frac{2q^2\hat E^2}{m\omega^2}\cos^2(kx) -
      \frac{\hbar q^2\hat E^2}{m^2c^2\omega}\sin\eta\,(\sin^2(kx)-\cos^2(kx))\,\sigma_x 
    \right)\Psi(x, t) \,,
  \end{equation}
  which is the searched-for evolution equation with a time-independent
  Hamiltonian.  It equals the Hamiltonian for the case of linear
  polarization \cite{Batelaan:2007:Colloquium_Illuminating} plus a term
  which accounts for the spin dynamics due to the ellipticity of the
  electromagnetic field.  Thus, we found polarization-dependent
  ponderomotive forces similar to the situation in
  Refs.~\cite{Pokrovsky:Kaplan:2005:Relativistic_reversal,Smorenburg:Kanters:etal:2011:Polarization-dependent_ponderomotive}.
  Note that the spin term originates in equal magnitude from both the
  Zeeman term $\sim\vec B\cdot\vec\sigma$ and the relativistic
  correction $\sim(\vec E\times\vec A)\cdot\vec\sigma$ (see also
  \cite{Bauke:Ahrens:etal:2014:Electron-spin_dynamics,Bauke:Ahrens:etal:2014:Electron-spin_dynamics_long}).

  Similarly, we can apply the Magnus expansion to
  \eqref{eq:Dirac_FW_equation_1d} with the vector potential of the
  antirotating case \eqref{eq:A_anti} and the corresponding
  electromagnetic fields, which yields now
  \begin{equation}
    \label{eq:U_anti}
    \Psi(x, t) = \exp\Bigg(
      -\frac{\I t}{\hbar}\Bigg(
      -\frac{\hbar^2}{2m}\frac{\partial^2}{\partial x^2} + 
      \frac{q^2 \hat E^2}{m\omega^2}\left(2\cos^2(kx+\eta)\,\cos\eta +1 - \cos\eta\right) 
      + \dots
      \Bigg) 
    \Bigg)\Psi(x, 0) \,,
  \end{equation}
  where we have again neglected in the exponent terms which do not grow
  linearly with time and/or are small in the Bragg regime.  Equation
  \eqref{eq:U_anti} represents the time-evolution operator of the
  equation
  \begin{equation}
    \label{eq:H_pond_anti_notyet}
    \I \hbar\dot \Psi(x, t)  = 
    \left( 
      -\frac{\hbar^2}{2m}\frac{\partial^2}{\partial x^2} + 
      \frac{q^2 \hat E^2}{m\omega^2}\left(2\cos^2(kx+\eta/2)\,\cos\eta +1 - \cos\eta\right) 
    \right)\Psi(x, t) \,.
  \end{equation}
\end{widetext}
This can be further simplified by a gauge transform, which removes the
additive constant, and a suitable shift of the coordinate system to
\begin{equation}
  \label{eq:H_pond_anti}
  \I \hbar\dot \Psi(x, t)  = 
  \left( 
    -\frac{\hbar^2}{2m}\frac{\partial^2}{\partial x^2} + 
    \frac{2q^2 \hat E^2}{m\omega^2}\cos^2(kx)\,\cos\eta 
  \right)\Psi(x, t) \,.
\end{equation}
In contrast to \eqref{eq:H_pond_co}, the Hamiltonian in
\eqref{eq:H_pond_anti} involves no spin-coupling terms, which is a
consequence of $\vec E\times \vec A=0$ and the fact that the Zeeman
interaction is zero on average (as it commutes with itself at different
times) in the antirotating case.

\section{Analytical solution of the relativistic Pauli equation and numerical results}
\label{sec:solution}

In the following, we will derive approximate analytical solutions of the
relativistic Pauli equations \eqref{eq:H_pond_co} and
\eqref{eq:H_pond_anti}.  For this purpose the wave function $\Psi(x, t)$
is expanded into plane waves as
\begin{equation}
  \label{eq:Pauli_momentum_eigen}
  \Psi(x, t) = \sqrt{\frac{k}{2 \pi}} \sum_{n=\ldots,-1,0,1,\ldots\atop \gamma\in \{\uparrow, \downarrow\}} 
  c_n^\gamma(t)\vec\chi^\gamma \e^{\I n k x} \,,
\end{equation}
with the time-dependent coefficients $c_n^\uparrow(t)$ and
$c_n^\downarrow(t)$, which will be conveniently combined into the pair
\begin{equation}
  c_n(t)=\trans{(c_n^{\uparrow}(t),c_n^{\downarrow}(t))} \,.
\end{equation}
The basis functions that are employed in \eqref{eq:Pauli_momentum_eigen}
represent states with a well-defined momentum and spin orientation.
They form a system of orthonormal functions on $x\in[0,\lambda]$.  Thus,
the evolution equation for $c_n^\gamma(t)$, the relativistic Pauli
equation in momentum space, can be derived by plugging the ansatz
\eqref{eq:Pauli_momentum_eigen} into \eqref{eq:H_pond_co} or
\eqref{eq:H_pond_anti} and taking the scalar product with $\sqrt{{k}/({2
    \pi})} \,\vec\chi^{\gamma'} \e^{\I n' k x}$.  We will ignore possible
turn-on and turn-off phases in the following, i.\,e., $\Delta T=0$, if
not otherwise identified.

\subsection{Corotating fields}
\label{sec:solution_co}

Using the expansion \eqref{eq:Pauli_momentum_eigen}, the relativistic
Pauli equation for corotating fields \eqref{eq:H_pond_co} is given in
momentum space by\allowdisplaybreaks[0]
\begin{multline}
  \label{eq:Pauli_FW_equation_momentum-space_co}
  \I \hbar\dot c_n(t) = 
  \frac{n^2k^2\hbar^2}{2 m}c_n(t) + 
  \frac{q^2{\hat E}^2}{2k^2 m c^2} (c_{n-2}(t)+2c_{n}(t)+c_{n+2}(t)) \\ +
  \frac{\hbar q^2 {\hat E}^2 \sin\eta}{2km^2 c^3}\sigma_x 
  (c_{n-2}(t)+c_{n+2}(t)) \,.
\end{multline}
\allowdisplaybreaks Odd and even modes are decoupled in
Eq.~\eqref{eq:Pauli_FW_equation_momentum-space_co}.  Furthermore, the
relation
\begin{equation}
  \label{eq:energy_relations}
  \frac{k^2\hbar^2}{2 m} \gg
  \frac{q^2{\hat E}^2}{2k^2 m c^2} >
  \frac{\hbar q^2 {\hat E}^2 \sin\eta}{2km^2 c^3}
\end{equation}
holds for the coefficients in
Eq.~\eqref{eq:Pauli_FW_equation_momentum-space_co} because of the
Bragg-regime condition $c|q|\hat E/\omega\ll \hbar\omega<mc^2$.  Thus,
it is justified to truncate the equation system, which yields for the
odd modes\pagebreak[1]
\begin{equation}
  \label{eq:matrix_co}
  \I
  \begin{pmatrix}
    \dot c_{-3}^\uparrow(t) \\[0.5ex]
    \dot c_{-3}^\downarrow(t) \\[0.5ex]
    \dot c_{-1}^\uparrow(t) \\[0.5ex]
    \dot c_{-1}^\downarrow(t) \\[0.5ex]
    \dot c_{1}^\uparrow(t) \\[0.5ex]
    \dot c_{1}^\downarrow(t) \\[0.5ex]
    \dot c_{3}^\uparrow(t) \\[0.5ex]
    \dot c_{3}^\downarrow(t) 
  \end{pmatrix} =
  \begin{pmatrix}
    9\Omega_1 & 0 & \Omega_2 & \Omega_3' & 0 & 0 & 0 & 0 \\
    0 & 9\Omega_1 & \Omega_3' & \Omega_2 & 0 & 0 & 0 & 0 \\
    \Omega_2 & \Omega_3' & \Omega_1 & 0 & \Omega_2 & \Omega_3' & 0 & 0 \\
    \Omega_3' & \Omega_2 & 0 & \Omega_1 & \Omega_3' & \Omega_2 & 0 & 0 \\
    0 & 0 & \Omega_2 & \Omega_3' & \Omega_1 & 0 & \Omega_2 & \Omega_3' \\
    0 & 0 & \Omega_3' & \Omega_2 & 0 & \Omega_1 & \Omega_3' & \Omega_2 \\
    0 & 0 & 0 & 0 & \Omega_2 & \Omega_3' & 9\Omega_1 & 0 \\
    0 & 0 & 0 & 0 & \Omega_3' & \Omega_2 & 0 & 9\Omega_1
  \end{pmatrix}
  \begin{pmatrix}
    c_{-3}^\uparrow(t) \\[0.5ex]
    c_{-3}^\downarrow(t) \\[0.5ex]
    c_{-1}^\uparrow(t) \\[0.5ex]
    c_{-1}^\downarrow(t) \\[0.5ex]
    c_{1}^\uparrow(t) \\[0.5ex]
    c_{1}^\downarrow(t) \\[0.5ex]
    c_{3}^\uparrow(t) \\[0.5ex]
    c_{3}^\downarrow(t) 
  \end{pmatrix}
\end{equation}
by introducing the frequencies
\begin{subequations}
  \begin{align}
    \Omega_1 & = \frac{k^2\hbar}{2 m} \,, \\
    \Omega_2 & = \frac{q^2{\hat E}^2}{2\hbar k^2 m c^2} \,, \\
    \Omega_3 & = \frac{q^2 {\hat E}^2}{2km^2 c^3} \,,
  \end{align}
\end{subequations}
and $\Omega_3'=\Omega_3\sin\eta$.  Note that an additive constant
$2\Omega_2$ has not been included on the diagonal of the coefficient
matrix in Eq.~\eqref{eq:matrix_co}, which can be justified by a suitable
gauge transform.  The eigenvalues and eigenvectors of this coefficient
matrix can be easily calculated by a computer algebra system giving some
rather intricate expressions.  In the parameter regime, where
\eqref{eq:energy_relations} is fulfilled, the eigenvectors of the
coefficient matrix can be approximated by the column vectors of the
constant matrix
\begin{multline}
  \label{eq:eigenvectors_co}
  \begin{pmatrix}
    \vec v_1 & \vec v_2 & \vec v_3 & \vec v_4 &
    \vec v_5 & \vec v_6 & \vec v_7 & \vec v_8
  \end{pmatrix} = \\
  \frac12
  \begin{pmatrix}
    0 & 0 & 0 & 0 & 1 & 1 & 1 & 1 \\
    0 & 0 & 0 & 0 & 1 & 1 & 1 & 1 \\
    1 & 1 & 1 & 1 & 0 & 0 & 0 & 0 \\
    1 & -1 & -1 & 1 & 0 & 0 & 0 & 0 \\
    1 & -1 & 1 & -1 & 0 & 0 & 0 & 0 \\
    1 & 1 & -1 & -1 & 0 & 0 & 0 & 0 \\
    0 & 0 & 0 & 0 & 1 & -1 & 1 & -1 \\
    0 & 0 & 0 & 0 & 1 & 1 & -1 & -1 
  \end{pmatrix}\,,
\end{multline}
and the corresponding exact eigenvalues are
\begin{subequations}
  \begin{multline}
    \varepsilon_{1,5} = 5\Omega_1 + \frac{\Omega_2+\Omega_3'}{2} \\ \mp
    \sqrt{{(8\Omega_1-(\Omega_2+\Omega_3'))^2}/4 + (\Omega_2+\Omega_3')^2} \,,
  \end{multline}\vspace*{-4ex minus 1ex}
  \begin{multline}
    \varepsilon_{2,6} = 5\Omega_1 + \frac{-\Omega_2+\Omega_3'}{2} \\ \mp
    \sqrt{{(8\Omega_1-(-\Omega_2+\Omega_3'))^2}/4 + (-\Omega_2+\Omega_3')^2} \,,
  \end{multline}\vspace*{-4ex minus 1ex}
  \begin{multline}
    \varepsilon_{3,7} = 5\Omega_1 + \frac{\Omega_2-\Omega_3'}{2} \\ \mp
    \sqrt{{(8\Omega_1-(\Omega_2-\Omega_3'))^2}/4 + (\Omega_2-\Omega_3')^2} \,,
  \end{multline}\vspace*{-4ex minus 1ex}
  \begin{multline}
    \varepsilon_{4,8} = 5\Omega_1 + \frac{-\Omega_2-\Omega_3'}{2} \\ \mp
    \sqrt{{(8\Omega_1-(-\Omega_2-\Omega_3'))^2}/4 + (-\Omega_2-\Omega_3')^2} \,.
  \end{multline}
\end{subequations}
Choosing $c_{-1}^\uparrow(0)=1$ and $c_{n}^\gamma(0)=0$ otherwise as the
initial condition and approximating the eigenvectors of the coefficient
matrix in Eq.~\eqref{eq:matrix_co} by \eqref{eq:eigenvectors_co} yields the
time-dependent solution to \eqref{eq:matrix_co},\pagebreak[1]
\allowdisplaybreaks[0]
\begin{subequations}
  \label{eq:sol_co}
  \begin{align}
    c^\uparrow_{-3}(t) & = 0 \,, \\
    c^\downarrow_{-3}(t) & = 0 \,, \\
    c^\uparrow_{-1}(t) & = \tfrac{1}{4}\left( 
      \e^{-\I\varepsilon_1 t} + \e^{-\I\varepsilon_2 t} + 
      \e^{-\I\varepsilon_3 t} + \e^{-\I\varepsilon_4 t}
    \right)\,, \\
    c^\downarrow_{-1}(t) & = \tfrac{1}{4}\left( 
      \e^{-\I\varepsilon_1 t} - \e^{-\I\varepsilon_2 t} - 
      \e^{-\I\varepsilon_3 t} + \e^{-\I\varepsilon_4 t}
    \right) \,, \\
    c^\uparrow_{1}(t) & = \tfrac{1}{4}\left(
      \e^{-\I\varepsilon_1 t} - \e^{-\I\varepsilon_2 t} + 
      \e^{-\I\varepsilon_3 t} - \e^{-\I\varepsilon_4 t}
    \right) \,, \\
    c^\downarrow_{1}(t) & = \tfrac{1}{4}\left(
      \e^{-\I\varepsilon_1 t} + \e^{-\I\varepsilon_2 t} -
      \e^{-\I\varepsilon_3 t} - \e^{-\I\varepsilon_4 t}
    \right) \,, \\
    c^\uparrow_{3}(t) & = 0 \,, \\
    c^\downarrow_{3}(t) & = 0 \,.
  \end{align}
\end{subequations}
\allowdisplaybreaks%
Note that although the amplitudes $c^{\uparrow/\downarrow}_{\pm3}(t)$
are zero, also including modes $n=\pm3$ in the truncated system
\eqref{eq:matrix_co} is crucial to describing the Kapitza-Dirac effect
properly for corotating fields.  The amplitudes
$c^{\uparrow/\downarrow}_{\pm3}(t)$ vanish only because the
approximation \eqref{eq:eigenvectors_co} has been applied.  Employing
the exact eigenvectors instead would yield small nonzero amplitudes.

\begin{figure*}
  \centering
  \includegraphics[scale=0.45]{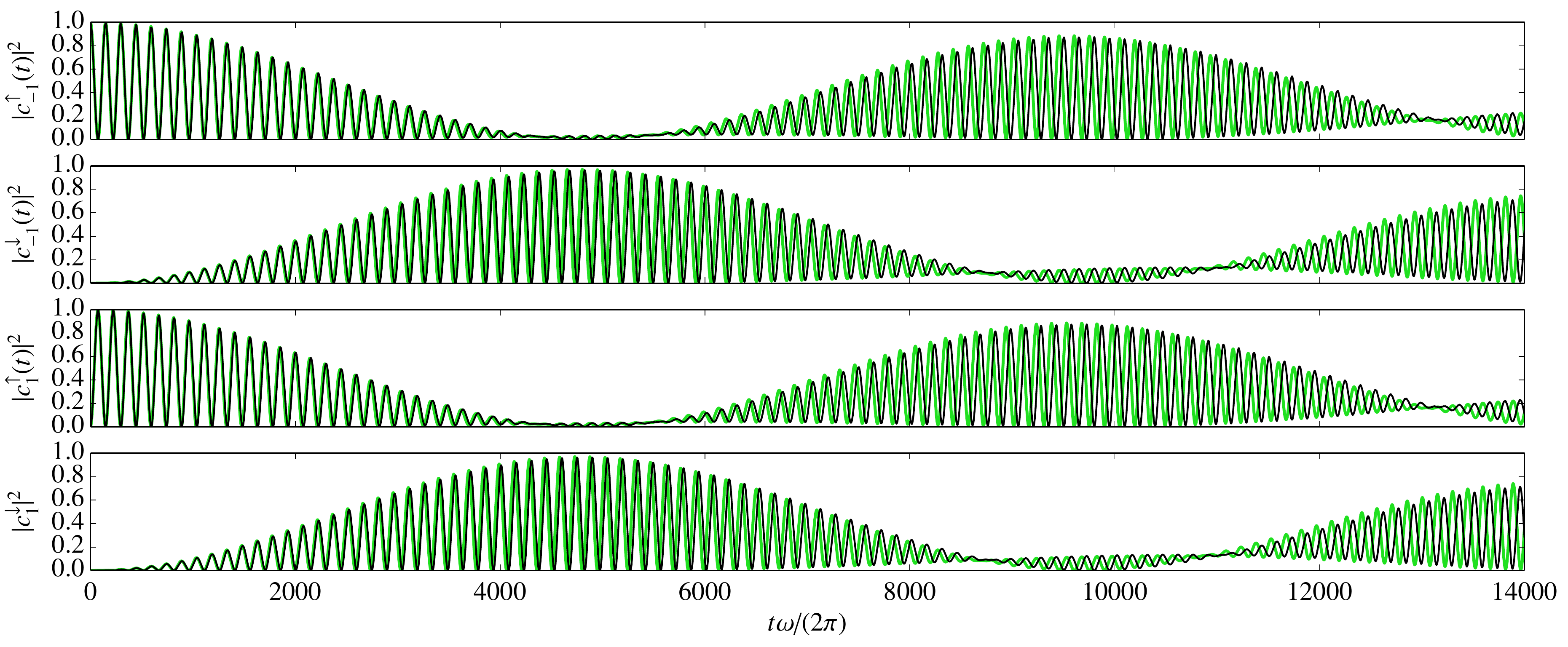}
  \caption{(Color online) Time evolution of the probabilities to find
    the electron, which interacts with a standing light wave of the
    corotating setup, in a particular quantum state.  Green (gray) lines
    show, from top to bottom, the analytical results \eqref{eq:c_t1_co}
    to \eqref{eq:c_t4_co} based on the relativistic Pauli equation
    \eqref{eq:H_pond_co}.  Black lines represent the corresponding
    numerical results based on the Dirac equation
    \eqref{eq:Dirac_equation}, i.\,e., the probabilities to find the
    electron in the states $\psi^{+\uparrow}_{-1}$,
    $\psi^{+\downarrow}_{-1}$, $\psi^{+\uparrow}_{1}$, and
    $\psi^{+\downarrow}_{1}$, respectively, after the electromagnetic
    field has been turned off.  Parameters are $\hat
    E=400\,\mathrm{a.u.}=2.06\times10^{14}\,\mathrm{V/m}$ (corresponding
    to an intensity of $1.12\times10^{22}\,\mathrm{W/cm^2}$),
    $\lambda=3\,\mathrm{a.u.}=0.159\,\mathrm{nm}$, $\Delta
    T=10\pi/\omega$, and $\eta=\pi/2$, corresponding to circular
    polarization.}
  \label{fig:corotating_prob}
\end{figure*}

The evolution of the probability $|c^\uparrow_{-1}(t)|^2$ is with
\eqref{eq:sol_co} \allowdisplaybreaks[0]
\begin{multline}
  \label{eq:c_up-12}
  |c^\uparrow_{-1}(t)|^2 =
  \tfrac{1}{16}\Big|
    \e^{-\I(\varepsilon_2+\varepsilon_3+\varepsilon_4)t} +
    \e^{-\I(\varepsilon_1+\varepsilon_3+\varepsilon_4)t} \\ +
    \e^{-\I(\varepsilon_1+\varepsilon_2+\varepsilon_4)t} +
    \e^{-\I(\varepsilon_1+\varepsilon_2+\varepsilon_3)t}
  \Big|^2\,.
\end{multline}
\allowdisplaybreaks
Considering the regime \eqref{eq:energy_relations}, the following
approximations hold:
\begin{subequations}
  \begin{align}
    \varepsilon_2+\varepsilon_3+\varepsilon_4 & \approx
    3\Omega_1 - \Omega_2 - \Omega_3' -
    \frac{3\Omega_2^2+3\Omega_3'^2-2\Omega_2\Omega_3'}{8\Omega_1}\,, \\
    \varepsilon_1+\varepsilon_3+\varepsilon_4 & \approx
    3\Omega_1 + \Omega_2 - \Omega_3' -
    \frac{3\Omega_2^2+3\Omega_3'^2+2\Omega_2\Omega_3'}{8\Omega_1}\,, \\
    \varepsilon_1+\varepsilon_2+\varepsilon_4 & \approx
    3\Omega_1 - \Omega_2 + \Omega_3' -
    \frac{3\Omega_2^2+3\Omega_3'^2+2\Omega_2\Omega_3'}{8\Omega_1}\,, \\
    \varepsilon_1+\varepsilon_2+\varepsilon_3 & \approx 
    3\Omega_1 + \Omega_2 + \Omega_3' -
    \frac{3\Omega_2^2+3\Omega_3'^2-2\Omega_2\Omega_3'}{8\Omega_1}\,.\!\!
  \end{align}
\end{subequations}
Employing these approximations in Eq.~\eqref{eq:c_up-12}, some algebraic
transformations yield
\begin{subequations}
  \label{eq:c_t_co}
  \begin{multline}
    \label{eq:c_t1_co}
    |c^\uparrow_{-1}(t)|^2 \approx 
    \cos^2(\Omega_2 t)\cos^2(\Omega_3't) \\ +
    (\sin^2(\Omega_2t)-\cos^2(\Omega_3't))
    \sin^2\left(\frac{\Omega_2\Omega_3't}{4\Omega_1}\right) \,.
  \end{multline}
  A similar calculation can be carried out for
  $|c^\downarrow_{-1}(t)|^2$, $|c^\uparrow_{1}(t)|^2$, and
  $|c^\downarrow_{1}(t)|^2$, which finally results in
  \allowdisplaybreaks[0]
  \begin{multline}
    \label{eq:c_t2_co}
    |c^\downarrow_{-1}(t)|^2 \approx
    \sin^2(\Omega_2 t)\sin^2(\Omega_3' t) \\ +
    (-\sin^2(\Omega_2t)+\cos^2(\Omega_3't))
    \sin^2\left(\frac{\Omega_2\Omega_3't}{4\Omega_1}\right) \,,
  \end{multline}\vspace*{-4ex}
  \begin{multline}
    \label{eq:c_t3_co}
    |c^\uparrow_{1}(t)|^2 \approx \sin^2(\Omega_2 t)\cos^2(\Omega_3' t) \\ +
    (-\sin^2(\Omega_2t)+\sin^2(\Omega_3't))
    \sin^2\left(\frac{\Omega_2\Omega_3't}{4\Omega_1}\right) \,,
  \end{multline}\vspace*{-4ex}
  \begin{multline}
    \label{eq:c_t4_co}
    |c^\downarrow_{1}(t)|^2 \approx \cos^2(\Omega_2 t)\sin^2(\Omega_3' t) \\ +
    (\sin^2(\Omega_2t)-\sin^2(\Omega_3't))
    \sin^2\left(\frac{\Omega_2\Omega_3't}{4\Omega_1}\right) \,.
  \end{multline}
\end{subequations}
\allowdisplaybreaks%
Thus, the occupation probabilities $|c^\uparrow_{-1}(t)|^2$,
$|c^\downarrow_{-1}(t)|^2$, $|c^\uparrow_{1}(t)|^2$, and
$|c^\downarrow_{1}(t)|^2$ show oscillatory behavior on different time
scales.  There is a fast frequency
\begin{equation}
  2\Omega_2 = \frac{q^2{\hat E}^2\lambda^2}{(2\pi)^2 \hbar m c^2}\,,
\end{equation}
an intermediate frequency 
\begin{equation}
  2\Omega_3' = \frac{q^2 {\hat E}^2 \lambda\sin\eta}{2\pi m^2 c^3} \,,
\end{equation}
and a slow frequency 
\begin{equation}
  \frac{\Omega_2\Omega_3'}{2\Omega_1} = 
  \frac{q^4 {\hat E}^4 \lambda^5\sin\eta}{(2\pi)^5\hbar^2m^2 c^5} \,.
\end{equation}
The fast oscillation's amplitude is modulated by the
inter\-mediate-frequency oscillation, and the fast oscillation's
amplitude and the intermediate oscillation's amplitude are modulated by
the slow-frequency oscillation.  The superposition of these three
amplitude-modulated oscillations creates a nontrivial temporal behavior
of the occupation probabilities (see also
Fig.~\ref{fig:corotating_prob}).

\begin{figure*}
  \centering
  \includegraphics[scale=0.45]{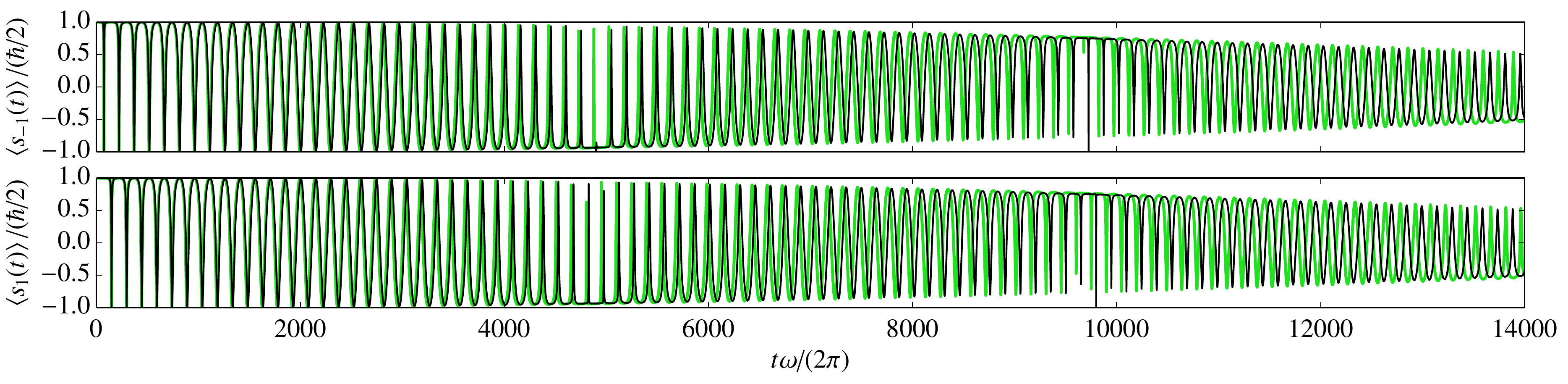}
  \caption{(Color online) Time evolution of the spin of the two
    diffraction modes.  Green (gray) lines show, from top to bottom, the
    analytical results \eqref{eq:spin_co_-1} and \eqref{eq:spin_co_1}
    based on the relativistic Pauli equation \eqref{eq:H_pond_co}.
    Black lines show the corresponding numerical results based on the
    Dirac equation \eqref{eq:Dirac_equation}.  Parameters are as in
    Fig.~\ref{fig:corotating_prob}.}
  \label{fig:corotating_spin}
\end{figure*}

For $\eta=0$, i.\,e., $\Omega_3'=0$, Eq.~\eqref{eq:c_t_co} simplifies to
the known result for the Kapitza-Dirac effect with linearly polarized
light \cite{Batelaan:2007:Colloquium_Illuminating}.  For nonzero
ellipticity, however, quantum transitions occur not only between states
with different momenta but also between states with different spin
orientations.  As a consequence, the standing wave's ellipticity induces
a beating behavior of the non-spin-resolved probabilities, i.\,e.,
\begin{multline}
  |c^\uparrow_{-1}(t)|^2+|c^\downarrow_{-1}(t)|^2 \approx
  \tfrac12(1+\cos(2\Omega_2t)\cos(2\Omega_3't)) \\
  =
  \tfrac12\cos^2((\Omega_2 + \Omega_3')t) +
  \tfrac12\cos^2((\Omega_2 - \Omega_3')t)\,,
\end{multline}\vspace*{-4ex}
\begin{multline}
  |c^\uparrow_{1}(t)|^2+|c^\downarrow_{1}(t)|^2\approx
  \tfrac12(1-\cos(2\Omega_2t)\cos(2\Omega_3't)) \\
  =
  \tfrac12\sin^2((\Omega_2 + \Omega_3')t) +
  \tfrac12\sin^2((\Omega_2 - \Omega_3')t)\,.
\end{multline}
With \eqref{eq:sol_co} the spin expectation value
\begin{equation}
  \braket{s(t)} = \frac{\hbar}{2}\left(
    |c^\uparrow_{-1}(t)|^2 - |c^\downarrow_{-1}(t)|^2 + 
    |c^\uparrow_{1}(t)|^2 - |c^\downarrow_{1}(t)|^2
  \right)
\end{equation}
is given by
\begin{equation}
  \braket{s(t)} = \frac{\hbar}{4}\left(
    \cos((\varepsilon_1-\varepsilon_3)t) +
    \cos((\varepsilon_2-\varepsilon_4)t) 
  \right) \,,
\end{equation}
which may be simplified to
\begin{equation}
  \label{eq:spin_co}
  \braket{s(t)} \approx \frac{\hbar}{2}
  \cos(2\Omega_3' t)\cos\left(\frac{\Omega_2\Omega_3'}{2\Omega_1}t\right)
\end{equation}
by employing 
\begin{subequations}
  \begin{align}
    \varepsilon_1-\varepsilon_3 & \approx
    2\Omega_3' - \frac{\Omega_2\Omega_3'}{2\Omega_1} \,, \\
    \varepsilon_2-\varepsilon_4 & \approx 2\Omega_3' +
    \frac{\Omega_2\Omega_3'}{2\Omega_1} \,.
  \end{align}
\end{subequations}
Thus, in contrast to the Kapitza-Dirac effect with linearly polarized
light, the electron's spin precesses in a standing wave of elliptical
polarization.  If the spin is conditioned to a specific mode, let us say
$n=-1$, one finds
\begin{multline}
  \label{eq:spin_co_-1}
  \braket{s_{-1}(t)} = 
  \frac{\hbar}{2}
  \frac{|c^\uparrow_{-1}(t)|^2 - |c^\downarrow_{-1}(t)|^2}
  {|c^\uparrow_{-1}(t)|^2 + |c^\downarrow_{-1}(t)|^2} \\ \approx
  \frac{\hbar}{2}\frac{
    \cos\left(\frac{\Omega_2\Omega_3'}{2\Omega_1}t\right)
    ( \cos(2\Omega_2 t) + \cos(2\Omega_3' t) )
  }{\cos(2\Omega_2 t)\cos(2\Omega_3' t) + 1} \,,
\end{multline}
and, similarly, for $n=1$ 
\begin{multline}
  \label{eq:spin_co_1}
  \braket{s_{1}(t)} = 
  \frac{\hbar}{2}
  \frac{|c^\uparrow_{1}(t)|^2 - |c^\downarrow_{1}(t)|^2}
  {|c^\uparrow_{1}(t)|^2 + |c^\downarrow_{1}(t)|^2} \\ \approx
  \frac{\hbar}{2}\frac{
    \cos\left(\frac{\Omega_2\Omega_3'}{2\Omega_1}t\right)
    ( \cos(2\Omega_2 t) - \cos(2\Omega_3' t) )
  }{\cos(2\Omega_2 t)\cos(2\Omega_3' t)-1} \,.
\end{multline}

\begin{figure}
  \centering
  \includegraphics[scale=0.45]{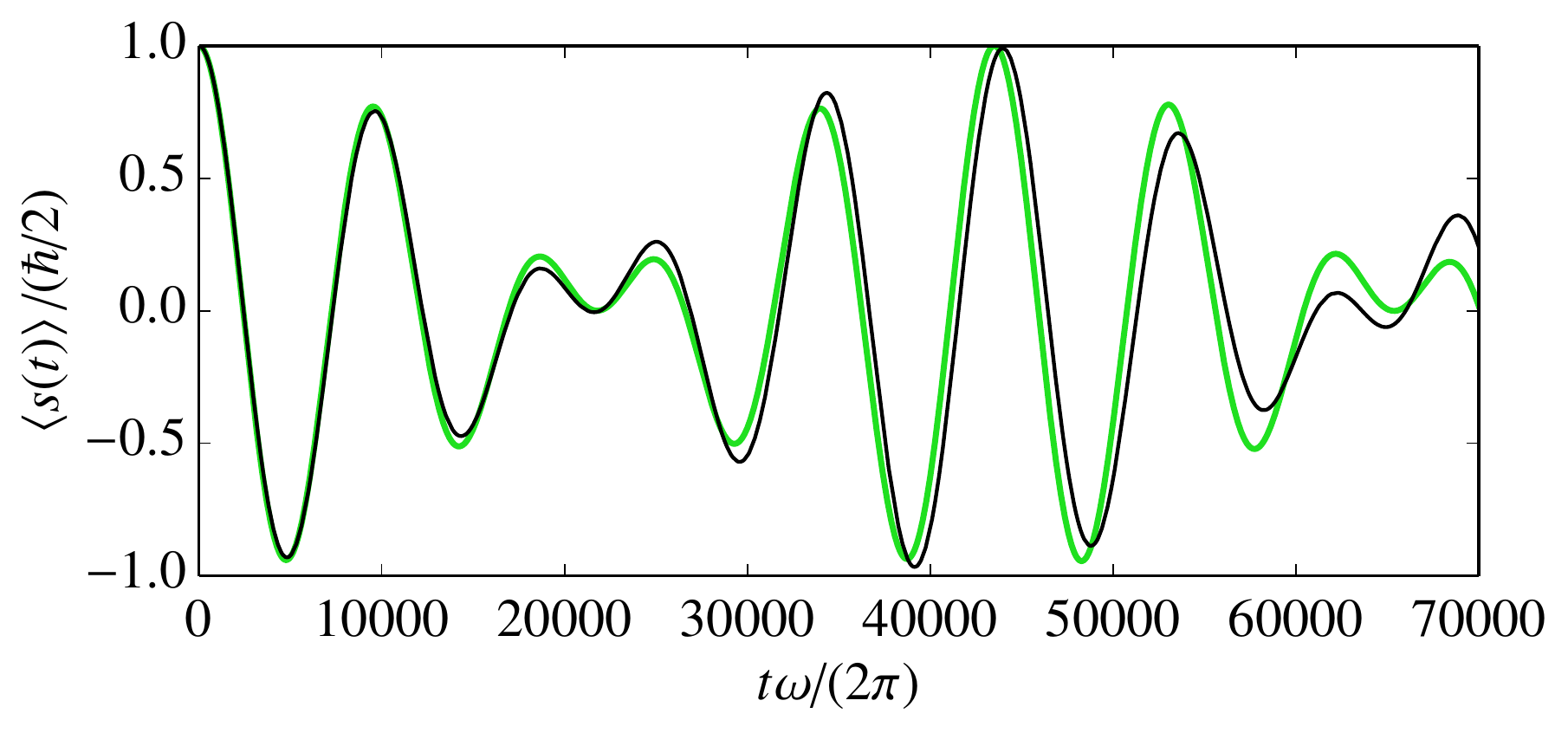}
  \caption{(Color online) Time evolution of the electron's spin.  Green
    (gray) line shows the analytical result \eqref{eq:spin_co} based on
    the relativistic Pauli equation \eqref{eq:H_pond_co}.  Black line
    shows the corresponding numerical data based on the Dirac equation
    \eqref{eq:Dirac_equation}.  Parameters are as in
    Fig.~\ref{fig:corotating_prob}.}
  \label{fig:corotating_total_spin}
\end{figure}

The analytical results, which have been obtained by deriving an
approximate solution to the relativistic Pauli equation
\eqref{eq:H_pond_co}, shall be compared to numerical solutions of the
time-dependent Dirac equation \eqref{eq:Dirac_equation} with space- and
time-dependent electromagnetic potentials.
Figure~\ref{fig:corotating_prob} presents the probabilities to find the
electron, which interacts with a standing light wave of the corotating
setup, in one of the quantum states $\psi^{+\uparrow}_{-1}$,
$\psi^{+\downarrow}_{-1}$, $\psi^{+\uparrow}_{1}$, and
$\psi^{+\downarrow}_{1}$ and compares this to the analytical predictions
\eqref{eq:c_t1_co} to \eqref{eq:c_t4_co}.  Even for rather long
interaction times the analytical results \eqref{eq:c_t1_co} to
\eqref{eq:c_t4_co} match the fully relativistic dynamics as predicted by
the Dirac equation.  For the time evolution of the expectation value of
the spin of the two diffraction modes, $\braket{s_{-1}(t)}$ and
$\braket{s_{1}(t)}$, as well as for the time evolution of the
expectation value of the spin $\braket{s(t)}$ we also find excellent
agreement between the Dirac theory and \eqref{eq:spin_co_-1},
\eqref{eq:spin_co_1}, and \eqref{eq:spin_co}, respectively (see
Figs.~\ref{fig:corotating_spin} and~\ref{fig:corotating_total_spin}).
The spin expectation values \eqref{eq:spin_co}, \eqref{eq:spin_co_-1},
and \eqref{eq:spin_co_1} oscillate on different time scales.  Because
$\Omega_2>\Omega_3'$, the shortest time scale of the total spin's
oscillation, which is determined by the frequency $2\Omega_3'$, is
longer than the time scale of the oscillation of \eqref{eq:spin_co_-1}
and \eqref{eq:spin_co_1}, which is determined by $2\Omega_2$.  Note that
the electron's spin precesses faster in the Kapitza-Dirac effect
compared to nonresonant scattering at the same field configuration,
where the spin precesses with the frequency
$\Omega_2\Omega_3'/(2\Omega_1)$ (see
Refs.~\cite{Bauke:Ahrens:etal:2014:Electron-spin_dynamics,Bauke:Ahrens:etal:2014:Electron-spin_dynamics_long}
and the \hyperref[eq:non_resonant]{Appendix}).

Although we employed in the presented numerical examples for the
Kapitza-Dirac dynamics ultra strong laser parameters in the soft
\mbox{x-ray} regime, the observed electron-spin dynamics is not a
strong-field effect.  Spin flips in the diffracted beam may be observed
also for less intense laser fields, but the spin-precession frequency
becomes small in this regime.  In this case the electron must be trapped
for many laser cycles in the laser field's focus to observe a full spin
flip.  The detectability of spin effects depends on the sensitivity of
the spin measurement and for how long the electron interacts with the
electromagnetic field.  If one requires that the electron shall have
reached full spin flip after $\mathcal{N}$ laser periods and the laser
parameters remain in the Bragg regime, then the required electric field
strength is bounded from below and from above as
\begin{equation}
  \label{eq:bounds_E}
  \sqrt{\frac{\omega^3\hbar m}{2q^2\mathcal{N}}} < \hat E \ll
  \frac{\omega^2\hbar}{|q|c} \,.
\end{equation}
Here the lower bound follows from $\mathcal{N}>\omega/(4\Omega_2)$, with
$2\Omega_2$ determining the shorter time scale of the spin dynamics of
the diffracted electron beam.  Note that the condition
\eqref{eq:bounds_E} is much less restrictive than the corresponding
relation for the nonresonant scattering as presented in
Ref.~\cite{Bauke:Ahrens:etal:2014:Electron-spin_dynamics_long}.

\begin{figure}
  \centering
  \includegraphics[scale=0.45]{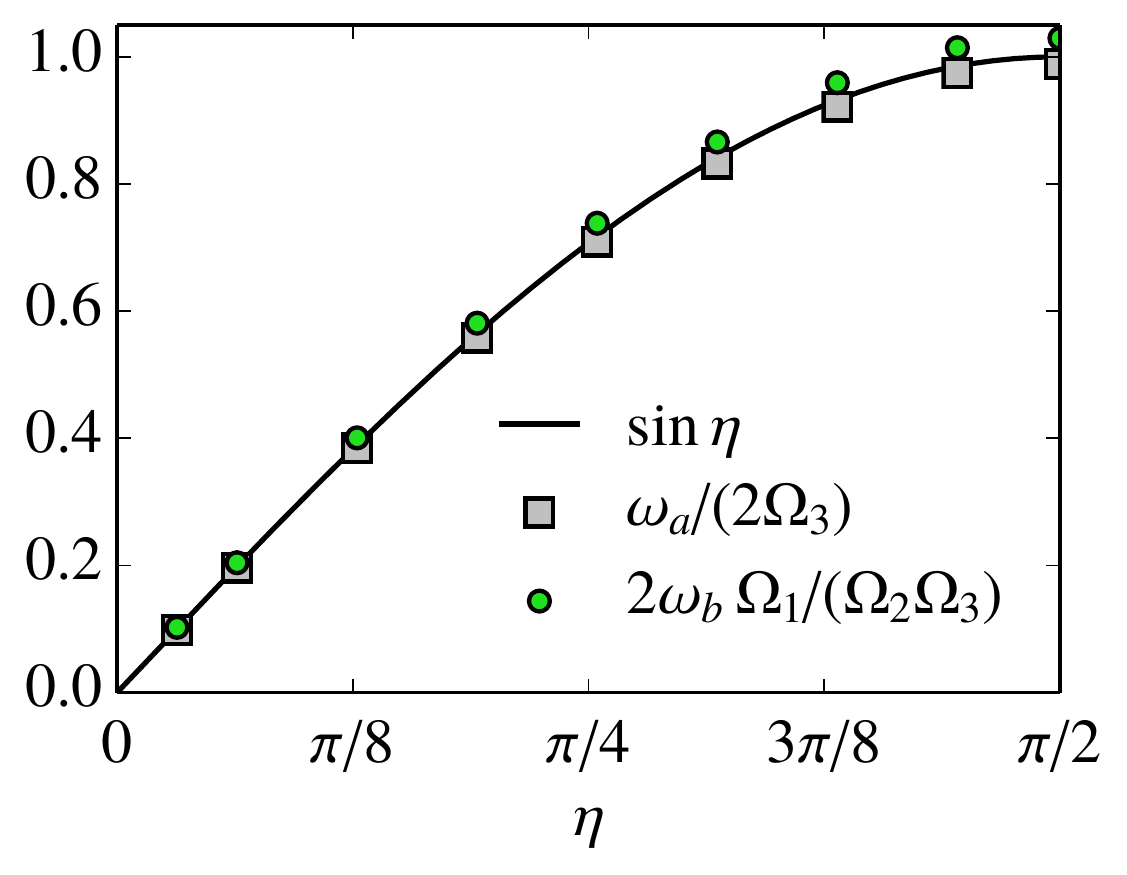}
  \caption{(Color online) Spin-oscillation frequencies for different
    degrees of ellipticity $\eta$ as obtained by a numerical fit
    procedure to data of numerical solutions of the Dirac equation
    compared to the predictions by the relativistic Pauli equation
    \eqref{eq:H_pond_co}. Parameters are as in
    Fig.~\ref{fig:corotating_prob}.}
  \label{fig:Omega_3}
\end{figure}

According to Eq.~\eqref{eq:spin_co} the electron's spin oscillates with
the frequency $2\Omega_3\sin\eta$, which is modulated by an oscillation
with the frequency $\Omega_2\Omega_3\sin\eta\,/(2\Omega_1)$.  This
allows us to test the effect of the light's ellipticity $\eta$ on the
Kapitza-Dirac effect.  For this purpose the numerical data for the spin
expectation value as a function of the interaction time were fitted to
$\hbar\cos(\omega_a t)\cos(\omega_b t)/2$ with the fit parameters
$\omega_a$ and $\omega_b$.  In Fig.~\ref{fig:Omega_3} the numerical
values of these fit parameters are compared to $2\Omega_3$ and
$\Omega_2\Omega_3/(2\Omega_1)$.  The numerical data show a clear
$\sin\eta$ dependency, as predicted by our analytical considerations.

\subsection{Antirotating fields}
\label{sec:solution_anti}

For the setup with antirotating fields, which is described by
\eqref{eq:H_pond_anti}, the ansatz \eqref{eq:Pauli_momentum_eigen} gives
the following set of equations in momentum space:
\begin{equation}
  \label{eq:Pauli_FW_equation_momentum-space_anti}
  \I \hbar\dot c_n(t) = 
  \frac{n^2k^2\hbar^2}{2 m}c_n(t) + 
  \frac{q^2{\hat E}^2\cos\eta}{2k^2 m c^2} (c_{n-2}(t)+2c_{n}(t)+c_{n+2}(t)) \,.
\end{equation}
Similar to the corotating case, only odd and even modes couple to
each other, and furthermore, modes corresponding to different spin
orientations are decoupled.  Truncating the system
\eqref{eq:Pauli_FW_equation_momentum-space_anti} to\pagebreak[1]
\begin{equation}
  \label{eq:matrix_anti}
  \I
  \begin{pmatrix}
    \dot c_{-1}^{\uparrow}(t) \\[0.5ex]
    \dot c_{-1}^{\downarrow}(t) \\[0.5ex]
    \dot c_{1}^{\uparrow}(t) \\[0.5ex]
    \dot c_{1}^{\downarrow}(t) 
  \end{pmatrix} =
  \begin{pmatrix}
    \Omega_1 & 0 & \Omega_2' & 0 \\
    0 & \Omega_1 & 0 & \Omega_2' \\
    \Omega_2' & 0 & \Omega_1 & 0 \\
     0 & \Omega_2' & 0 & \Omega_1
  \end{pmatrix}
  \begin{pmatrix}
    c_{-1}^{\uparrow}(t) \\[0.5ex]
    c_{-1}^{\downarrow}(t) \\[0.5ex]
    c_{1}^{\uparrow}(t) \\[0.5ex]
    c_{1}^{\downarrow}(t) 
  \end{pmatrix}
\end{equation}
with $\Omega_2'=\Omega_2\cos\eta$ yields, for the initial condition
$c_{-1}^\uparrow(0)=1$ and $c_{n}^\gamma(0)=0$ otherwise, the exact
solution
\begin{subequations}
  \begin{align}
    c^\uparrow_{-1}(t) & = 
    \cos(\Omega_1t)\cos(\Omega_2't) 
    -\I\sin(\Omega_1t)\cos(\Omega_2't) \,, \\
    c^\downarrow_{-1}(t) & = 0 \,, \\
    c^\uparrow_{1}(t) & =
    -\sin(\Omega_1t)\sin(\Omega_2') 
    - \I\cos(\Omega_1t)\sin(\Omega_2') \,, \\
    c^\downarrow_{1}(t) & = 0 \,.
  \end{align}
\end{subequations}
The probabilities $|c^\uparrow_{-1}(t)|^2$ and
$|c^\uparrow_{1}(t)|^2$ follow as
\begin{subequations}
  \begin{align}
    |c^\uparrow_{-1}(t)|^2 & = \cos^2(\Omega_2't) \,, \\
    |c^\uparrow_{1}(t)|^2 & = \sin^2(\Omega_2't) \,.
  \end{align}
\end{subequations}
Thus, the probabilities oscillate with the Rabi frequency
\begin{equation}
  \label{eq:Rabi_freq_anti}
  2\Omega_2' = \frac{q^2{\hat E}^2\lambda^2\cos\eta}{(2\pi)^2\hbar m c^2} \,,
\end{equation}
which is known as the Rabi frequency of the Kapitza-Dirac effect with
linearly polarized light, modified by the factor $\cos\eta$.

\begin{figure}[b]
  \centering
  \includegraphics[scale=0.45]{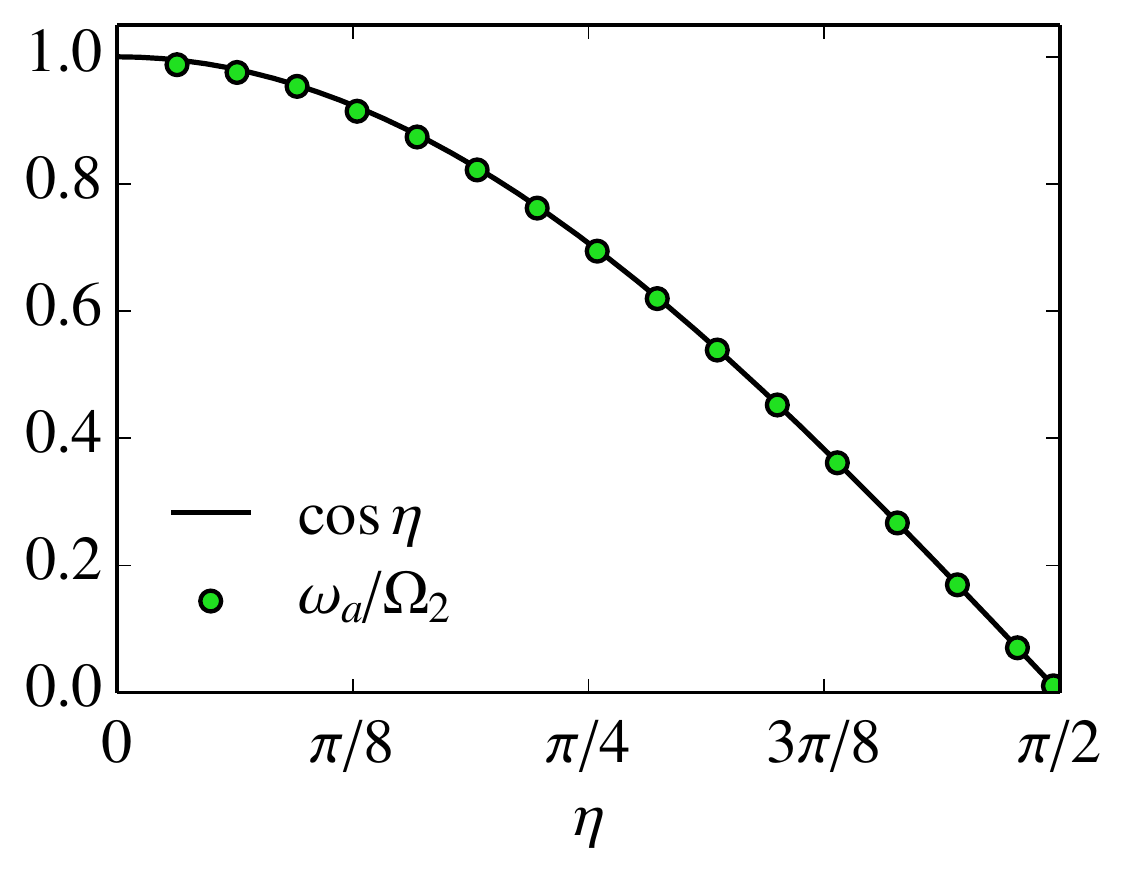}
  \caption{(Color online) Rabi oscillation frequencies for different
    degrees of ellipticity $\eta$ for the antirotating setup as obtained
    by a numerical fit procedure to data of numerical solutions of the
    Dirac equation compared to the predictions by the relativistic Pauli
    equation \eqref{eq:H_pond_co}. Parameters are as in
    Fig.~\ref{fig:corotating_prob}.}
  \label{fig:Omega_2}
\end{figure}

As a consequence of \eqref{eq:Rabi_freq_anti}, a nonvanishing
ellipticity decreases the Rabi frequency.  In the case of circular
polarization it even becomes zero; that is, Kapitza-Dirac scattering is
completely suppressed.  To test this analytical prediction we fitted the
probability to find the electron in the quantum state
$\psi^\uparrow_{-1}$ as obtained from the numerical solution of the
Dirac equation to the function $\cos^2(\omega_at)$ with the fit
parameter $\omega_a$.  The numerical values for $\omega_a$ are compared
to $\Omega_2'$ for different values of $\eta$ in Fig.~\ref{fig:Omega_2},
which shows a clear $\cos\eta$ dependency, as predicted by our analytical
considerations.

As demonstrated above, the dynamics of the electron in the antirotating
setup can be modeled by a time-independent scalar potential, which may
be understood by classical arguments.  Consider a charged particle in
the antirotating fields.  The classical motion of the particle is
determined by the Lorentz force
\begin{align}
  \ddot{\vec{r}} = 
  \frac{q}{m} \left( 
    \vec E\antirot(x, t) +
    \dot{\vec{r}} \times \vec B\antirot(x, t) 
  \right) \,.
\end{align}
The velocity $\dot{\vec{r}}$ may be divided into two parts,
$\dot{\vec{r}} = \dot{\vec{r}}_{\parallel} + \dot{\vec{r}}_{\perp}$,
that are parallel to the $x$~axis and perpendicular to the $x$~axis,
respectively.  The change of the latter is primarily determined by the
electrical field
\begin{equation}
  \ddot{\vec{r}}_{\perp} = \frac{q}{m} \vec E\antirot(x, t) \,.
\end{equation}
Integrating this equation of motion yields the perpendicular velocity
\begin{equation}
  \dot{\vec{r}}_{\perp} =
  \frac{2 q \hat{E}}{m \omega} \sin(\omega t)
  \begin{pmatrix}
    0 \\ \cos(kx) \\ \cos(kx + \eta)
  \end{pmatrix}
  + \dot{\vec{r}}_{\perp}(0) \,.
\end{equation}
The motion along the $x$~axis is solely determined by the magnetic
field, i.\,e.,
\begin{multline} 
  \label{eq:r..para}
  \ddot{\vec{r}}_{\parallel} = 
  \frac{q}{m} \dot{\vec{r}}_{\perp} \times \vec B\antirot(x, t) \\ =
  \frac{4 q^{2} \hat{E}^{2} \sin^{2}(\omega t)}{m^{2} c \omega}
  \begin{pmatrix}
    0 \\ \cos(kx) \\ \cos(kx + \eta)
  \end{pmatrix}\times
  \begin{pmatrix}
    0 \\ -\sin(kx + \eta) \\ \sin(kx)
  \end{pmatrix}\\
  + \frac{q}{m} \dot{\vec{r}}_{\perp}(0) \times \vec B\antirot(x, t) \,
  \vec{e}_{x} \,.
\end{multline}
Because the velocity $\dot{\vec{r}}_{\perp}$ is in phase with the
magnetic field and both are proportional to $\sin(\omega t)$, the
acceleration parallel to the $x$~axis does not average out over a laser
cycle.  Only the effect of the last term in Eq.~\eqref{eq:r..para}
cancels for sufficiently long interaction times.  The value of the
nonvanishing contribution depends crucially on the spatial phase
relation between the electric field, which is $\sim\trans{(0, \cos(kx),
  \cos(kx+\eta))}$, and the magnetic field, which is $\sim\trans{(0,
  -\sin(kx+\eta), \sin(kx))}$.  Thus, the time-averaged acceleration
parallel to the $x$~axis
\begin{align}
  \ddot{\vec{r}}_{\parallel} \approx 
  \frac{2 q^{2} \hat{E}^{2}}{m^{2} c \omega } \cos\eta\,\sin(2kx+\eta) \, \vec{e}_{x}
\end{align}
and the corresponding potential 
\begin{align}
  V_\mathrm{pond} = 
  \frac{2 q^{2} \hat{E}^{2}}{m \omega^{2}} \cos\eta\,\cos^{2}(kx+\eta/2)
\end{align}
are proportional to $\cos\eta$, which causes the $\cos\eta$ dependency
of the Rabi frequency \eqref{eq:Rabi_freq_anti}.

\section{Conclusions}
\label{sec:conclusions}

We studied the two-photon Kapitza-Dirac effect in the Bragg regime for
setups with counterpropagating elliptically polarized electromagnetic
waves whose electric-field components may have the same or an opposite
sense of rotation.  The time-dependent fully relativistic Dirac equation
was solved numerically to simulate the electron's dynamics.  To allow
for an analytical description of the scattering process the explicitly
time-dependent Dirac-equation Hamiltonian was reduced to an effective
time-independent eight-level Hamiltonian or four-level Hamiltonian by
approximating the Dirac equation by the Pauli equation plus the leading
relativistic corrections and employing suitable ponderomotive
potentials.  In the parameter range of the Bragg regime, the fully
relativistic Dirac equation and the effective time-independent
Hamiltonian give qualitatively the same results, which also agree
quantitatively very well for not too long interaction times.

The light waves' ellipticity does not change the Bragg condition
compared to the case of linear polarization.  It induces, however, spin
dynamics in the case of corotating counterpropagating waves.  The spin's
expectation value of the quantum-mechanical superposition of the
scattered and nonscattered states precesses with a period which is
larger than the Rabi period of the Kapitza-Dirac scattering.  The
projections of the electron's quantum state onto the scattering and
nonscattering channels, however, show spin precession on the time scale
of the Rabi period of Kapitza-Dirac scattering.  The precession of the
electron's spin is induced equally by the Zeeman interaction and the
electron-spin's coupling to the photonic spin density of the standing
light wave.  This effect may be observable by employing upcoming laser
sources for circularly polarized light
\cite{Depresseux:Oliva:etal:2015:Demonstration_of} with wavelength and
intensity parameters as utilized by Freimund
et~al.~\cite{Freimund:Aflatooni:Batelaan:2001:Observation_of}, provided
that the electron interacts long enough with the electromagnetic field
to rotate the spin.  Compared to nonresonant scattering
\cite{Bauke:Ahrens:etal:2014:Electron-spin_dynamics_long,Bauke:Ahrens:etal:2014:Electron-spin_dynamics},
where spin precession is also induced by the elliptically polarized
light wave, the spin-precession frequency is substantially faster for
the Kapitza-Dirac effect.  Thus, shorter interaction times are required
to observe spin-precession experimentally.

Antirotating counterpropagating waves yield a standing wave with linear
polarization, i.\,e., vanishing photonic spin density; thus, no spin
effects can be observed for this kind of setup.  However, the phase
between the electric and the magnetic-field components of the standing
wave depends on the ellipticity of the counterpropagating laser fields,
and in this way the Rabi period of Kapitza-Dirac scattering depends on
the ellipticity.  This Rabi period diverges in the limit of circular
polarization; that is, the Kapitza-Dirac effect cannot be observed in
the antirotating setup for circular polarization.

\appendix*

\section{Nonresonant scattering}
\label{eq:non_resonant}

Recently, it was shown
\cite{Bauke:Ahrens:etal:2014:Electron-spin_dynamics_long} that the
coupling of the spin angular momentum of light beams with elliptical
polarization to the spin degree of freedom of free electrons can lead to
spin precession.  The spin-precession frequency was derived by a tedious
calculation based on time-dependent perturbation theory.  The setup that
was considered in
Refs.~\cite{Bauke:Ahrens:etal:2014:Electron-spin_dynamics_long,Bauke:Ahrens:etal:2014:Electron-spin_dynamics}
is identical to the corotating field configuration, which was0 analyzed
in Sec.~\ref{sec:solution_co}, except that the electron is initially at
rest; that is, it does not fulfill the Bragg condition of the
Kapitza-Dirac effect.  In the following we will show that the
spin-precession frequency of
\cite{Bauke:Ahrens:etal:2014:Electron-spin_dynamics_long} can be derived
easily by employing the time-independent Hamiltonian in
Eq.~\eqref{eq:H_pond_co}.

Equation \eqref{eq:Pauli_FW_equation_momentum-space_co} yields, for the
even modes, the truncated system\pagebreak[1]
\begin{equation}
  \label{eq:matrix_co_bar}
  \I
  \begin{pmatrix}
    \dot c_{-2}^\uparrow(t) \\[0.5ex]
    \dot c_{-2}^\downarrow(t) \\[0.5ex]
    \dot c_{0}^\uparrow(t) \\[0.5ex]
    \dot c_{0}^\downarrow(t) \\[0.5ex]
    \dot c_{2}^\uparrow(t) \\[0.5ex]
    \dot c_{2}^\downarrow(t) 
  \end{pmatrix} =
  \begin{pmatrix}
    4\Omega_1 & 0 & \Omega_2 & \Omega_3' & 0 & 0 \\
    0 & 4\Omega_1 & \Omega_3' & \Omega_2 & 0 & 0 \\
    \Omega_2 & \Omega_3' & 0 & 0 & \Omega_2 & \Omega_3' \\
    \Omega_3' & \Omega_2 & 0 & 0 & \Omega_3' & \Omega_2 \\
    0 & 0 & \Omega_2 & \Omega_3' & 4\Omega_1 & 0 \\
    0 & 0 & \Omega_3' & \Omega_2 & 0 & 4\Omega_1
  \end{pmatrix}
  \begin{pmatrix}
    c_{-2}^\uparrow(t) \\[0.5ex]
    c_{-2}^\downarrow(t) \\[0.5ex]
    c_{0}^\uparrow(t) \\[0.5ex]
    c_{0}^\downarrow(t) \\[0.5ex]
    c_{2}^\uparrow(t) \\[0.5ex]
    c_{2}^\downarrow(t) 
  \end{pmatrix} \,.
\end{equation}
As for the Kapitza-Dirac effect the eigenvalues and the eigenvectors of
the coefficient matrix in Eq.~\eqref{eq:matrix_co_bar} can be calculated
exactly, which gives the exact eigenvalues 
\begin{subequations}
  \begin{align}
    \varepsilon_{1,2} & = 
    2\Omega_1 - 
    \sqrt{4\Omega_1^2 + 2(\Omega_2\pm\Omega_3')^2} \,, \\
    \varepsilon_{3,4} & = 4 \Omega_1 \,, \\
    \varepsilon_{5,6} & = 
    2\Omega_1 + 
    \sqrt{4\Omega_1^2 + 2(\Omega_2\pm\Omega_3')^2} \,.
  \end{align}
\end{subequations}
However, we replace the exact eigenvectors by the approximate
eigenvectors\pagebreak[1]
\begin{equation}
  \label{eq:eigenvectors_co_bar}
  \begin{pmatrix}
    \vec v_1 & \vec v_2 & \vec v_3 & \vec v_4 &
    \vec v_5 & \vec v_6 
  \end{pmatrix} = 
  \begin{pmatrix}
    0 & 0 & \frac{1}{\sqrt 2} & 0 & \frac12 & \frac12 \\[0.75ex]
    0 & 0 & 0 & \frac{1}{\sqrt 2} & \frac12 & -\frac12 \\[0.75ex]
    \frac{1}{\sqrt 2} & \frac{1}{\sqrt 2} & 0 & 0 & 0 & 0 \\[0.75ex]
    \frac{1}{\sqrt 2} & -\frac{1}{\sqrt 2} & 0 & 0 & 0 & 0 \\[0.75ex]
    0 & 0 & -\frac{1}{\sqrt 2} & 0 & \frac12 & \frac12 \\[0.75ex]
    0 & 0 & 0 & -\frac{1}{\sqrt 2} & \frac12 & -\frac12
  \end{pmatrix}\,.
\end{equation}
Specifying $c_{0}^\uparrow(0)=1$ and $c_{n}^\gamma(0)=0$ otherwise as
the initial condition and approximating the eigenvectors of the
coefficient matrix in Eq.~\eqref{eq:matrix_co_bar} by
\eqref{eq:eigenvectors_co_bar} yields the time-dependent solution to
\eqref{eq:matrix_co_bar},
\begin{subequations}
  \label{eq:sol_co_bar}
  \begin{align}
    c^\uparrow_{2}(t) & = 0 \,, \\
    c^\downarrow_{2}(t) & = 0 \,, \\
    c^\uparrow_{0}(t) & = \tfrac12\left(
      \e^{-\I\varepsilon_1 t} + \e^{-\I\varepsilon_2 t} 
    \right)\,, \\
    c^\downarrow_{0}(t) & =  \tfrac12\left(
      \e^{-\I\varepsilon_1 t} - \e^{-\I\varepsilon_2 t} 
    \right)\,, \\
    c^\uparrow_{2}(t) & = 0 \,, \\
    c^\downarrow_{2}(t) & = 0 \,.
  \end{align}
\end{subequations}
The evolution of the probabilities $|c^\uparrow_{0}(t)|^2$ and
$|c^\downarrow_{0}(t)|^2$ is with \eqref{eq:sol_co_bar}\pagebreak[1]
\begin{subequations}
  \begin{align}
    |c^\uparrow_{0}(t)|^2 & = \cos^2\left(
      \frac{\varepsilon_2-\varepsilon_1}{2}t
    \right)\,, \\
    |c^\downarrow_{0}(t)|^2 & = \sin^2\left(
      \frac{\varepsilon_2-\varepsilon_1}{2}t 
    \right)\,.
  \end{align}
\end{subequations}
Considering again the regime \eqref{eq:energy_relations}, the 
approximation
\begin{equation}
  \varepsilon_2-\varepsilon_1 \approx
  \frac{2\Omega_2\Omega_3'}{\Omega_1}
\end{equation}
holds.  Thus,\pagebreak[1]
\begin{subequations}
  \begin{align}
    |c^\uparrow_{0}(t)|^2 & \approx \cos^2\left(
      \frac{q^4 \hat E^4 \lambda^5\sin\eta}{(2\pi)^5\hbar^2m^2c^5}\frac{t}{2}
    \right)\,, \\
    |c^\downarrow_{0}(t)|^2 & \approx \sin^2\left(
      \frac{q^4 \hat E^4 \lambda^5\sin\eta}{(2\pi)^5\hbar^2m^2c^5}\frac{t}{2}
    \right)\,,
  \end{align}
\end{subequations}
which represents exactly the Rabi oscillation of the electron spin as
predicted in
Refs.~\cite{Bauke:Ahrens:etal:2014:Electron-spin_dynamics_long,Bauke:Ahrens:etal:2014:Electron-spin_dynamics}.

\bibliography{Kapitza_Dirac_ellipical}

\end{document}